\documentclass[11pt,a4paper]{article}
\usepackage{tabularx}
\usepackage[dvipdfm]{graphicx}
\usepackage{jheppub}

\def\numu{$\nu_{\mu}$ }
\def\nue{$\nu_{e}$ }
\def\nutau{$\nu_{\tau}$ }
\def\dm2{$\Delta m^{2}$ }

\def\eV2{eV$^{2}$}
\def\sinsqtt13{sin$^{2}(2\theta _{13})$}
\def\sin22t{sin$^{2}(2\theta)$}
\def\sin22tnew{sin$^{2}(2\theta _{new})$}

\title{Search for $\nu_\mu \rightarrow \nu_e$ oscillations with the OPERA experiment in the CNGS beam}

\collaboration{The OPERA collaboration}
\author[a]{N.~Agafonova,}
\author[b]{A.~Aleksandrov,}
\author[c]{A.~Anokhina,}
\author[d]{S.~Aoki,}
\author[e,*]{A.~Ariga\note[*]{Corresponding authors.},}
\author[e]{T.~Ariga,}
\author[f]{D.~Autiero,}
\author[g]{A.~Badertscher,}
\author[e]{A.~Ben Dhahbi,}
\author[h]{A.~Bertolin,}
% ...

\affiliation[a]{INR-Institute for Nuclear Research of the Russian Academy of Sciences, RUS-117312 Moscow, Russia}
\affiliation[b]{INFN Sezione di Napoli, I-80125 Napoli, Italy}
\affiliation[c]{(MSU SINP) Lomonosov Moscow State University Skobeltsyn Institute of Nuclear Physics, RUS-119992 Moscow, Russia}
\affiliation[d]{Kobe University, J-657-8501 Kobe, Japan}
\affiliation[e]{Albert Einstein Center for Fundamental Physics, Laboratory for High Energy Physics (LHEP), University of Bern, CH-3012 Bern, Switzerland}
\affiliation[f]{IPNL, Universit\'e Claude Bernard Lyon I, CNRS/IN2P3, F-69622 Villeurbanne, France}
\affiliation[g]{ETH Zurich, Institute for Particle Physics, CH-8093 Zurich, Switzerland}
\affiliation[h]{INFN Sezione di Padova, I-35131 Padova, Italy}
% ...

\emailAdd{akitaka.ariga@lhep.unibe.ch}
\emailAdd{umut.kose@cern.ch}

\abstract{
A first result of the search for \numu $\rightarrow$ \nue oscillations in the OPERA experiment, located at the Gran Sasso Underground Laboratory, is presented. The experiment looked for the appearance of \nue in the CNGS neutrino beam using the data collected in 2008 and 2009. 
Data are compatible with the non-oscillation hypothesis in the three-flavour mixing model. 
A further analysis of the same data constrains the non-standard oscillation parameters $\theta_{new}$ and $\Delta m^2_{new}$ suggested by the LSND and MiniBooNE experiments. 
For large $\Delta m^{2}_{new}$ values ($>$0.1 eV$^{2}$), the OPERA 90\% C.L. upper limit on \sin22tnew based on a Bayesian statistical method reaches the value $7.2 \times 10^{-3}$.
}

\dedicated{This paper is dedicated to the memory of Prof. Perihan Tolun.}

\begin{document}
\maketitle
\flushbottom

\section{Introduction}

   The OPERA experiment \cite{proposal} is designed to perform an appearance search for the \numu $\rightarrow$ \nutau oscillations \cite{PMNS} in the CNGS \numu beam \cite{CNGS} produced at CERN and directed towards the OPERA detector at the Gran Sasso Underground Laboratory (LNGS), 730 km away.
A charged-current (CC) \nutau interaction in the lead-emulsion target can be identified by detecting the decay of the short-lived $\tau$ lepton through particle tracking in the high-resolution nuclear emulsions. 
The observation of two \nutau candidate events has recently been reported \cite{firsttau,run0809,neutrino2012}.
The tracking capabilities of emulsions also allow to identify electrons produced in CC interactions of \nue and therefore to search for \nue appearance from \numu $\rightarrow$ \nue oscillations.
Given the long baseline of the experiment, and the high energy of the \numu beam ($\langle E\rangle$ = 17 GeV), OPERA has a good sensitivity for $\Delta m^{2}>0.01$ eV$^{2}$, i.e. for the LSND \cite{lsnd} - MiniBooNE \cite{miniboone} allowed region (see section \ref{sec:2flavor}). Recently ICARUS, which shares the beamline with OPERA, severely limited this region \cite{icarus}. Here we present a further constraint on these non-standard oscillations
from the analysis of the data collected in 2008 and 2009.

\section{Detector, beam and data taking}

In OPERA, neutrinos interact in a large mass target made of lead plates interspaced with nuclear emulsion films acting as high accuracy tracking devices \cite{Detector, emulsion}. 
This kind of target is historically called Emulsion Cloud Chamber (ECC). 
The full OPERA detector is made of a veto plane followed by two identical Super Modules (SM), each consisting of a target section and a magnetic muon spectrometer. The target sections are made of, in total, 150,000 emulsion/lead ECC modules (or "bricks") arranged in planes, with a weight of about 1,250 tons, interleaved by the scintillator "Target Tracker" (TT) planes. A target brick consists of 56 1-mm thick lead plates interleaved with 57 emulsion films for a total weight of 8.3 kg. 
Its thickness along the beam direction corresponds to about 10 X$_{0}$, which is optimized to detect $\nu_\mu \rightarrow \nu_\tau$ oscillations.
Tightly packed removable doublets of emulsion films, called Changeable Sheets (CS) \cite{cs}, are placed on the downstream face of each brick. 
They serve as interfaces between the TT planes and the bricks to facilitate the location of the neutrino interactions. 

Charged particles from a neutrino interaction in a brick cross the CS and produce signals in the scintillator strips of the TT. 
These signals are used to trigger the read-out and identify the brick where the interaction occurred. 
The brick is then extracted by an automated system. After development, the emulsion films are sent to the scanning laboratories.

\begin{figure}[htbp]
	\center
	\includegraphics[width=11cm]{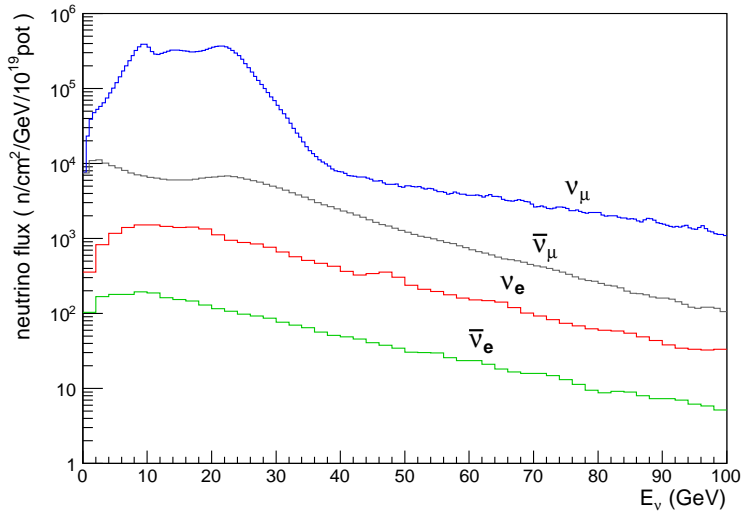}
	\caption{Neutrino fluxes of the different components at Gran Sasso in log scale.}
	\label{fig:flux}
	\end{figure}

The CNGS \numu beam, to which the OPERA detector is exposed, contains a small contamination of $\overline{\nu}_\mu$, \nue, and $\overline{\nu}_e$. 
The energy spectra at the detector, as obtained from a Monte Carlo simulation \cite{CNGSFlux}, are shown in figure \ref{fig:flux}. The integrated contamination of \nue and $\overline{\nu}_{e}$ CC interactions at Gran Sasso, relative to the integrated number of \numu CC interactions, is 0.88\% and 0.05\%, respectively.

OPERA collected data corresponding to $17.97 \times 10^{19}$ protons on target (pot) by December 2012 with 18941 events recorded.
The analysis reported in this paper uses the data collected in 2008 and 2009, corresponding to $5.25 \times10^{19}$ pot ($1.73 \times10^{19}$ and $3.52 \times 10^{19}$ pot, respectively) and to 5255 events recorded. The details of data taking and a comparison with Monte Carlo (MC) simulations for the 2008 and 2009 runs are reported in \cite{run0809, ED}.

\section{Emulsion scanning and search for \nue interactions}

    Bricks that are candidates for containing neutrino interactions are analysed following a complex procedure described in detail in \cite{firsttau,run0809}. 
Here we just recall the main steps of the analysis.

   The TT predictions are used for a large area scan of the corresponding CS films. 
If candidate tracks corresponding to the TT predictions are found in the CS, the 57 films contained in the brick are developed, and sent to the scanning labs. 
The tracks found in the CS are then followed upstream from film to film (scan-back) to find the neutrino interaction vertex. 
Once the vertex is found, the scanning of a volume downstream of the vertex (1 cm$^{2}$ in area and at least 7 films or 1.2 X$_0$ in thickness) is performed in order to reconstruct all the tracks connected to the vertex and to search for decay topologies.

\begin{figure}[htbp]
	\center
	\includegraphics[width=12cm,height=4cm]{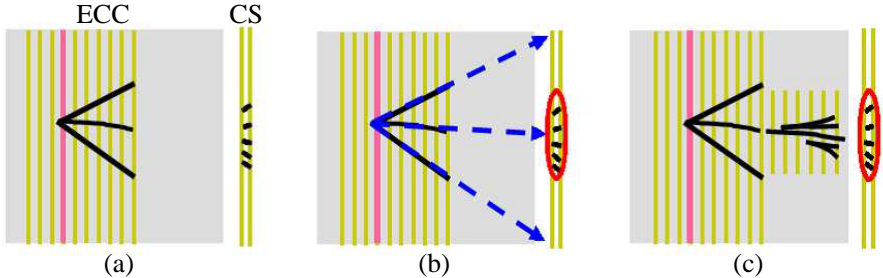}
	\caption{Sketch for the procedure of a systematic search for \nue candidates. After the reconstruction of tracks in the standard volume (a), all tracks emerging from the interaction vertex (the pink film) are extrapolated to the CS (b). If 3 or more tracks are found in the CS, corresponding to a given track, an additional volume along the full track length is scanned, leading to the detection of the electromagnetic shower (c).}
	\label{fig:syssearch}
	\end{figure}

  The main goal is the observation of the decay of a $\tau$ lepton, as a signature for a \nutau CC interaction. Moreover, the scanning also allows identifying electrons, hence \nue CC interactions. 
The identification of an electron is essentially based on the detection of the associated electromagnetic shower. 
Since the size of the standard scanned volume is too short in the beam direction to contain the electromagnetic shower, the search for electrons is performed using an extended scanning volume defined by a dedicated procedure sketched in figure \ref{fig:syssearch}.
 All primary tracks emerging from the interaction vertex are extrapolated to the CS. The tracks with angles similar ($\Delta \theta <150$ mrad) to that of the corresponding primary track (figure \ref{fig:syssearch}b) are searched in the CS region within 2 mm around the projected point.
If 3 or more tracks are found in the CS, corresponding to a given primary track, an additional volume along the candidate track is scanned, aiming at the reconstruction of an electromagnetic shower (figure \ref{fig:syssearch}c).

\begin{figure}[htbp]
	\center
	\includegraphics[width=9cm,height=5cm]{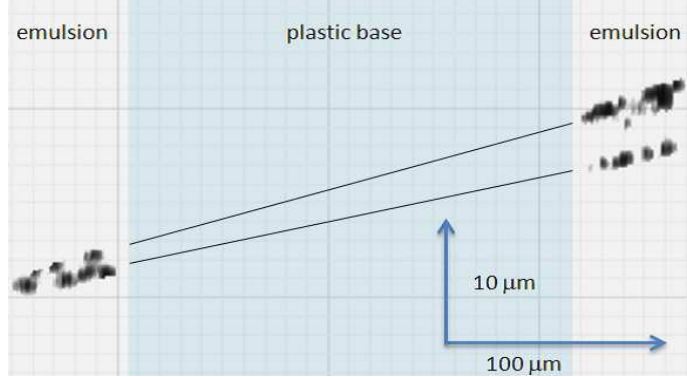}
	\caption{Side view of an $e^{+}e^{-}$ pair detected in an emulsion film. Note, that in a film the two emulsion layers are separated by a 205 $\mu$m thick plastic base. }
	\label{fig:epair}
	\end{figure}

   If a shower is found, the corresponding primary track becomes an electron candidate. 
The candidate track is then carefully inspected in the first two emulsion films following the interaction vertex. 
The aim is to check whether the track is due to a single particle (an electron) or to an $e^{+}e^{-}$ pair and so to reject electromagnetic showers initiated by the early conversion of a $\gamma$ from a $\pi^0$ decay.
Figure \ref{fig:epair} shows, as an example, the reconstruction of an $e^{+}e^{-}$ pair from a $\gamma$ conversion in the two layers of one emulsion film. 
The figure illustrates the capability to measure tracks with micrometric resolution. 
To remove $\gamma$ background, those configurations are excluded where a track can be separated into two almost parallel segments more than 1 $\mu$m apart in the first or second film.

A significant impact parameter of the electron track with respect to the primary vertex would allow to identify the event as a \nutau CC interaction with a $\tau \rightarrow e$ decay. For the present analysis, an upper limit of 10 $\mu$m is set on this impact parameter to select tracks originating from the vertex.

In addition, a scan-back procedure along the electromagnetic shower, described in \cite{proposal}, is applied. Implemented to increase the detection efficiency for a $\tau \rightarrow e$ decay in the \nutau quasi-elastic interactions, it is also beneficial for the detection of \nue CC interactions.

Once the presence of an electron track is confirmed at the neutrino interaction vertex, the event is classified as a \nue interaction. 
The energy of the \nue candidates is estimated from the reconstructed energy deposition in the TT making use of a calibration obtained through the MC simulation in a similar way as for the \numu CC and neutral-current (NC) events as described in \cite{ED}. The estimated energy resolution for an energy range up to 100 GeV can be parametrized as:
\begin{equation}
\nonumber
\Delta E/E = 0.37+0.74/\sqrt{E}\;\; (E\; \textrm{in GeV}).
\end{equation}

\begin{figure}[htbp]
	\center
	\includegraphics[width=11cm]{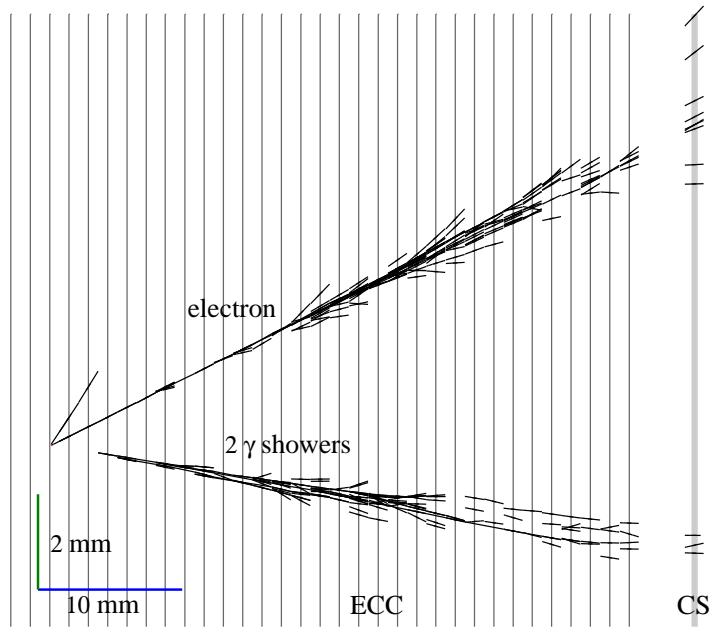}
	\caption{Display of the reconstructed emulsion tracks of one of the \nue candidate events. The reconstructed neutrino energy is 32.5 GeV. Two tracks are observed at the neutrino interaction vertex. One of the two generates an electromagnetic shower and is identified as an electron. In addition, two showers from $\gamma$ conversions are observed (overlapping in this projection), starting from 2 and 3 films downstream of the vertex.}
	\label{fig:candidate}
	\end{figure}

Among the 5255 candidate neutrino interactions collected during the 2008 and 2009 runs, 2853 vertices were localized in the bricks, out of which 505 did not have a muon identified by the electronic detectors, i.e. were not classified as \numu CC interactions. 
Out of those 505 events 19 \nue candidate events were found ; 17 events were found with the procedure illustrated in figure \ref{fig:syssearch}, and the 2 remaining events were found with the scan-back procedure mentioned before.
To illustrate the typical pattern of \nue candidates, figure \ref{fig:candidate} shows the reconstructed image of a \nue candidate event, with the track segments observed along the showering electron track.

\begin{figure}[htbp]
	\center
	\includegraphics[width=10cm]{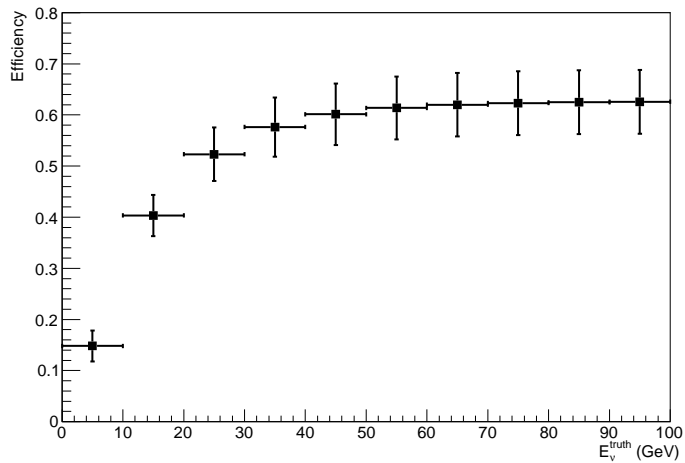}
	\caption{Detection efficiency of \nue events as a function of the neutrino energy, obtained from MC simulations. The error bars show the estimated systematic uncertainties.}
	\label{fig:eff}
	\end{figure}

The \nue detection efficiency as a function of the neutrino energy is computed with a GEANT3 based MC simulation. The simulated events are reconstructed with the same algorithms as used for the data. The results of the simulation are shown in figure \ref{fig:eff}.
The efficiency drop at low energy is strongly related to the smaller number of hits in the TT and the CS, due to the absorption of electromagnetic showers in the bricks. 
In order to optimize the performance of the experiment, the scanning and analysis strategies were tuned along the years. All strategies are tested in the MC and the deviations are taken as systematic uncertainties. 
Larger deviations are observed at lower energy, depending on the number of tracks observed in the CS. For events with energy above 10 GeV, the systematic uncertainty is estimated to be 10\% and 20\% below. 
Averaged over its energy spectrum, the \nue beam contamination detection efficiency is $\varepsilon_{det}=(53\pm 5)\%$.
For neutrino energies smaller than 30 GeV (20 GeV) it is $\varepsilon_{det} = (43\pm 5)\%\; ((35\pm 4)\%)$.

Apart from the \nue background associated to the beam contamination (see section \ref{sec:oscillation}), two main sources of background are considered for the \nue search: (a) $\pi^0$ misidentified as electron in neutrino interactions without a reconstructed muon; (b) \nutau CC interactions with the decay of the $\tau$ into an electron. 

Background (a) occurs if an $e^{+}e^{-}$ pair appears to be connected to the interaction vertex and cannot be distinguished from a single particle in the first two emulsion films after the vertex or if one branch of the pair has a very low energy and remains undetected. This background was evaluated directly from the data. 
In 1106 neutrino interactions, $\gamma$s converting in the second and third lead plates after the interaction vertex were searched for, and the above described procedure was applied to them; 1 event passed the criteria for the \nue search.
This result was converted into the probability to observe background \nue candidates due to $\gamma$ conversions in the first lead plate, taking into account the radiation length.
By normalizing to our sample, the estimated background (a) is $0.2\pm 0.2$ events. This number is compatible with an independent MC evaluation. The effect of $\pi^{0}$ decays to a Dalitz pair is estimated to be one order of magnitude smaller than the above value; therefore it is neglected.

 Background (b) was computed by MC simulation assuming the three-flavour $\nu_\mu \rightarrow \nu_\tau$ oscillation at maximal mixing and $\Delta m^{2}=2.32\times 10^{-3}$ eV$^{2}$ \cite{PDG}.
This background comes mainly from $\tau$ decaying in the same lead plate as the primary vertex with the impact parameter of the daughter electron to the primary vertex smaller than 10 $\mu$m; secondarily from an undetected kink ($\theta_{kink}<20$ mrad) from $\tau$ decaying in further downstream material. Background (b) is estimated to be $0.3\pm 0.1$ events in our sample.

The total amount of the considered background for the \nue CC interaction search is $0.4\pm 0.2$ events.

\section{Oscillation analysis}
\label{sec:oscillation}

\subsection{Background to $\nu_\mu \rightarrow \nu_e$ appearance}

   A signal for \numu $\rightarrow$ \nue oscillations should appear as a significant excess of electron events with respect to the expected background, mainly due to \nue and $\overline{\nu}_{e}$ CC interactions from the beam contamination.  A detailed evaluation of this number was performed, starting from the fluxes of the different beam components presented in figure \ref{fig:flux}.
The simulation of the neutrino fluxes and spectra including a description of all beam line elements is based on the FLUKA MC code \cite{fluka1,fluka2}. Details on the simulation can be found in \cite{beamcernnote}.
Conservatively a 10\% systematic uncertainty on the \nue beam contamination has been considered as in \cite{icarus}.
However it is worth noting that this number affects marginally the sensitivity of the measurement which is dominated by the small sample size. 
The fluxes were weighted with the CC cross sections and the energy dependent detection efficiency. An additional inefficiency of 6\% (for 2008) and 3\% (for 2009) was introduced to reflect the lower film quality of a small fraction of the bricks. Taking into account the target mass and the pot corresponding to our data, we expect to observe $19.4\pm 2.8\; (syst)$ \nue events from the beam contamination in the full energy range. Together with the backgrounds (a) and (b) discussed above, we expect $19.8\pm 2.8\; (syst)$ background \nue events. This number is in agreement with the 19 observed candidate \nue events and therefore the room for oscillations is reduced. In the following we analyse two scenarios.

\subsection{Three-flavour mixing scenario}

A non-zero $\theta_{13}$ has recently been reported by several experiments \cite{dayabay,reno,wchooz,t2k}. Using the following oscillation parameters \cite{PDG} : sin$^{2}(2\theta_{13})=0.098$, sin$^2(2\theta_{23}) =1$, $\Delta m^{2}_{32} = \Delta m^{2}_{31} = 2.32\times 10^{-3}$ eV$^{2}$, also assuming $\delta_{CP}=0$ and neglecting matter effects, 1.4 oscillated \nue CC events are expected to be detected in the whole energy range.

\begin{figure}[htbp]
	\center
	\includegraphics[width=11cm]{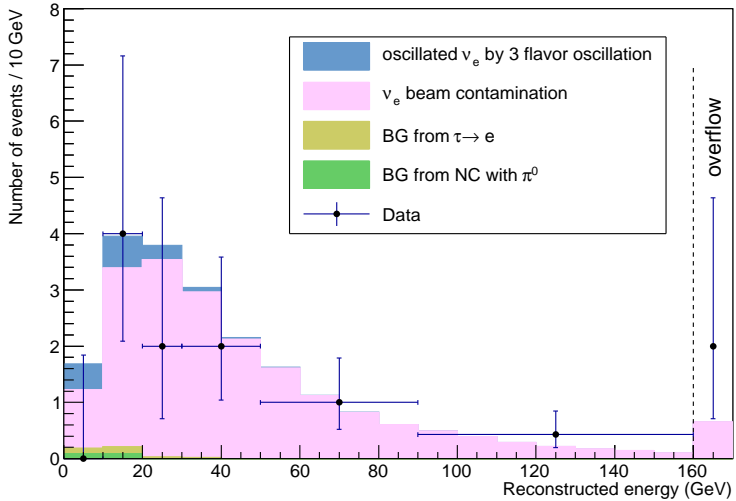}
	\caption{Distribution of the reconstructed energy of the \nue events, and the expected spectrum from the different sources in a stack histogram, normalized to the number of pot analysed for this paper. 
	}
	\label{fig:energy}
	\end{figure}

Figure \ref{fig:energy} shows the reconstructed energy distribution of the 19 \nue candidates, compared with the expected reconstructed energy spectra from the \nue beam contamination, the oscillated \nue from the three-flavour oscillation and the background (a) and (b), normalized to the pot analysed for this paper. To increase the signal to background ratio a cut $E < 20$ GeV is applied on the reconstructed energy of the event, which provides the best figure of merit on the sensitivity to $\theta_{13}$. Within this cut, 4.2 events from \nue beam contamination and 0.4 events from the backgrounds (a) and (b) are expected, while 4 events are observed. The numbers are summarized in table \ref{tab:numbers}.
The number of observed events is compatible with the non-oscillation hypothesis and an upper limit \sinsqtt13$ < 0.44$ is derived at the 90\% Confidence Level (C.L.).

\begin{table}
\small
\begin{tabular}{|l|l|r|r|r|}
\hline
\multicolumn{2}{|l|}{Energy cut } & 20 GeV & 30 GeV & No cut  \\
\hline \hline
BG common to    & BG (a) from $\pi ^{0}$            & 0.2  &  0.2  &    0.2    \\ \cline{2-5}
both analyses   & BG (b) from $\tau \rightarrow e$  & 0.2  &  0.3  &    0.3    \\ \cline{2-5}
                & \nue beam contamination           & 4.2  &  7.7  &   19.4    \\ \hline
\multicolumn{2}{|l|}{Total expected BG in 3-flavour oscillation analysis} 
                                                    & 4.6  &  8.2  &   19.8  \\ \hline \hline
BG to non-standard        & \nue via 3-flavour oscillation 
                                                    &  1.0 &  1.3  &    1.4   \\ 
oscillation analysis only &                                &       &       &        \\ \hline
\multicolumn{2}{|l|}{Total expected BG in non-standard oscillation analysis} 
                                                    & 5.6  &  9.4  &   21.3  \\ 
\hline
\hline
\multicolumn{2}{|l|}{Data}                          & 4    &  6    &   19    \\
\hline

\end{tabular}
\caption{Expected and observed number of events for the different energy cuts.}
\label{tab:numbers}
\end{table}

\subsection{Non-standard oscillations}
\label{sec:2flavor}

   Beyond the three-neutrino paradigm, some possible hints for non-standard effects have been reported, in particular by the LSND \cite{lsnd} and MiniBooNE \cite{miniboone} experiments. We have used OPERA data to set an upper limit on non-standard \numu $\rightarrow$ \nue oscillations. 

We used the conventional approach of expressing the $\nu_\mu \rightarrow \nu_e$ oscillation probability in the one mass scale dominance approximation, given by the following formula with new oscillation parameters $\theta_{new}$ and $\Delta m^2_{new}$ :
\begin{equation}
\nonumber
P_{\nu_\mu \rightarrow \nu_e}=\textrm{sin}^{2}(2\theta _{new})\cdot \textrm{sin}^{2}(1.27\Delta m^2_{new}L(\textrm{km})/E(\textrm{GeV}))\end{equation}
Note however that this approach does not allow a direct comparison between experiments working in different L/E regimes \cite{Bhattacharya:2011ee}.

The \numu flux at the detector, normalized to the integrated statistics used in our analysis, is weighted by the oscillation probability, by the CC cross-section and by the energy dependent detection efficiency, to obtain the number of \nue CC events expected from this oscillation.

As the energy spectrum of the oscillated \nue with large $\Delta m^{2}_{new}$ ($>$0.1 eV$^{2}$) follows the spectrum of $\nu_{\mu}$, which is basically vanishing above 40 GeV (see figure \ref{fig:flux}), a cut on the reconstructed energy is introduced. 
The optimal cut on the reconstructed energy in terms of sensitivity is found to be 30 GeV. We observe 6 events below 30 GeV (69\% of the oscillation signal at large $\Delta m^{2}_{new}$ is estimated to remain in this region), while the expected number of events from background is estimated to be $9.4\pm 1.3\; (syst)$ (see table \ref{tab:numbers}). Note that we choose to include the three-flavour oscillation induced events into the background. In this case, the oscillation probability does not contain the $\theta_{13}$ driven term.

The 90\% C.L. upper limit on \sin22tnew is then computed by comparing the expectation from oscillation plus backgrounds, with the observed number of events. 
Since we observed a smaller number of events than the expected background, we provide both, the Feldman and Cousins (F\&C) confidence intervals \cite{FC} and the Bayesian bounds, setting a prior to zero in the unphysical region and to a constant in the physical region \cite{cowan}. Uncertainties of the background were incorporated using prescriptions provided in \cite{PDG}. The results obtained from the two methods for the different C.L. are reported in table \ref{tab:ul}. We also quote our sensitivity calculated assuming 9 observed events (integer number closest to the expected background).

\begin{table}
\center
\begin{tabular}{|l|r|r|r|r|r|}
\hline
       &      &\multicolumn{2}{c|}{Upper limit} & \multicolumn{2}{c|}{Sensitivity}\\ \cline{3-6}
       & C.L. & \multicolumn{1}{c|}{F\&C} & \multicolumn{1}{c|}{Bayes} & \multicolumn{1}{c|}{F\&C} & \multicolumn{1}{c|}{Bayes} \\ 
\hline
Number of oscillated       & 90\% & 3.1 & 4.5 & 6.1 & 6.5 \\
\nue events     & 95\% & 4.3 & 5.7 & 7.8 & 7.9 \\
              & 99\% & 6.7 & 8.2 & 10.7 & 10.9 \\
\hline
$\textrm{sin}^{2}(2\theta _{new})$ at                          & 90\% & 5.0$\times10^{-3}$ & 7.2$\times10^{-3}$ & 9.7$\times10^{-3}$ & 10.4$\times10^{-3}$ \\
large \dm2  & 95\% & 6.9$\times10^{-3}$ & 9.1$\times10^{-3}$ & 12.4$\times10^{-3}$ & 12.7$\times10^{-3}$ \\
            & 99\% & 10.6$\times10^{-3}$ & 13.1$\times10^{-3}$ & 17.1$\times10^{-3}$ & 17.4$\times10^{-3}$ \\
\hline
\end{tabular}
\caption{Upper limits on the number of oscillated \nue CC events and $\textrm{sin}^{2}(2\theta _{new})$, obtained by the F\&C and Bayesian methods, for C.L. 90\%, 95\%, 99\%. The sensitivity is computed assuming that the number of observed events is 9, which is the closest integer to the 9.4 expected background events.}
\label{tab:ul}

\end{table}

\begin{figure}[htbp]
	\center
	\includegraphics[width=12cm]{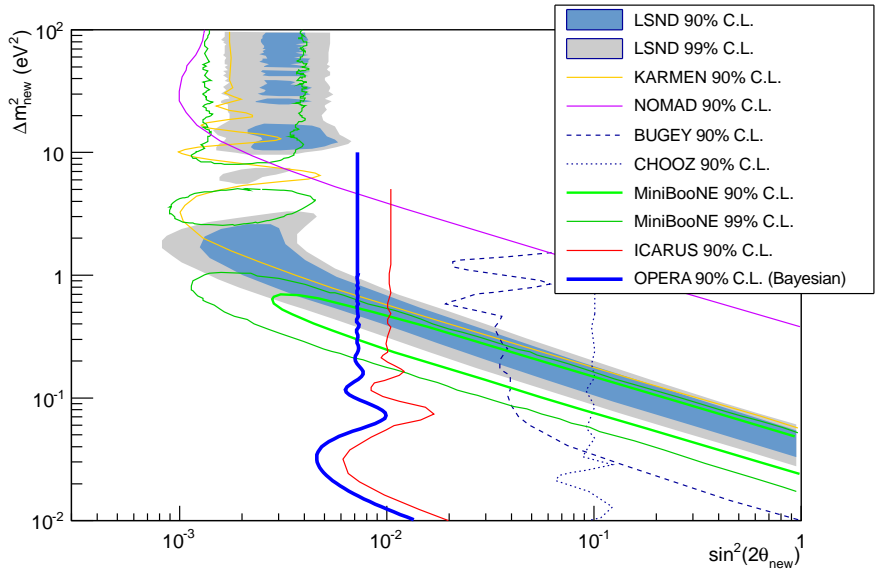}
	\caption{The exclusion plot for the parameters of the non-standard $\nu_\mu \rightarrow \nu_e$ oscillation, obtained from this analysis using the Bayesian method, is shown. The other limits shown, mostly using frequentist methods, are from KARMEN ($\overline{\nu}_\mu \rightarrow \overline{\nu}_{e}$ \cite{karmen}), BUGEY ($\overline{\nu}_{e}$ disappearance \cite{bugey}), CHOOZ ($\overline{\nu}_{e}$ disappearance \cite{chooz}), NOMAD ($\nu_{\mu} \rightarrow \nu_{e}$ \cite{nomad}) and ICARUS ($\nu_{\mu} \rightarrow \nu_{e}$ \cite{icarus}). 
The regions corresponding to the positive indications reported by  LSND ($\overline{\nu}_{\mu} \rightarrow \overline{\nu}_{e}$ \cite{lsnd}) and MiniBooNE  (\numu $\rightarrow$ \nue and $\overline{\nu}_{\mu} \rightarrow \overline{\nu}_{e}$ \cite{miniboone}) are also shown.}
	\label{fig:exclusion}
	\end{figure}

Given the underfluctuation of the data, the curve with the Bayesian upper limit was chosen for the exclusion plot shown in figure \ref{fig:exclusion}. For convenience, results from the other experiments, working at different $L/E$ regimes, are also reported in this figure.
For large $\Delta m^{2}_{new}$ values the OPERA 90\% upper limit on \sin22tnew reaches the value $7.2 \times 10^{-3}$, while the sensitivity corresponding to the pot used for this analysis is $10.4 \times 10^{-3}$.

As seen in figure \ref{fig:energy}, the underfluctuation is mainly present in the low energy region. In order to illustrate the impact of energy cuts on our analysis, in table \ref{tab:ulenergycuts}, the limits are quoted for different cuts.

\begin{table}
\center
\begin{tabular}{|l|r|r|r|r|}
\hline
             &\multicolumn{2}{c|}{Upper limit} & \multicolumn{2}{c|}{Sensitivity}\\ \cline{2-5}
  Energy cut & \multicolumn{1}{c|}{F\&C} & \multicolumn{1}{c|}{Bayes} & \multicolumn{1}{c|}{F\&C} & \multicolumn{1}{c|}{Bayes} \\ 
\hline
20 GeV        &  8.5$\times10^{-3}$ & 10.4$\times10^{-3}$ & 14.2$\times10^{-3}$ & 14.2$\times10^{-3}$ \\
30 GeV        &  5.0$\times10^{-3}$ & 7.2$\times10^{-3}$ & 9.7$\times10^{-3}$ & 10.4$\times10^{-3}$ \\
No cut        &  8.6$\times10^{-3}$ & 9.5$\times10^{-3}$ & 10.8$\times10^{-3}$ & 11.0$\times10^{-3}$ \\
\hline
\end{tabular}
\caption{90\% C.L. upper limits and sensitivities on $\textrm{sin}^{2}(2\theta _{new})$, for different energy cuts, according to the F\&C and Bayesian methods.}
\label{tab:ulenergycuts}
\end{table}

\section{Conclusions and perspectives}

First results of a search for \numu $\rightarrow$ \nue oscillations with the OPERA experiment at the Gran Sasso Underground Laboratory have been presented. The experiment searched for the appearance of \nue in the CNGS neutrino beam using the data collected in 2008 and 2009, corresponding to an integrated intensity of 5.25 $\times$ 10$^{19}$ pot. The observation of 19 \nue candidate events is compatible with the non-oscillation expectation of 19.8$\pm$2.8 events.

The current result on the search for the three-flavour neutrino oscillation yields an upper limit \sinsqtt13 $<$ 0.44 (90\% C.L.).

OPERA limits the parameter space available for a non-standard \nue appearance suggested by the results of the LSND and MiniBooNE experiments. It further constrains the still allowed region around $\Delta m^{2}_{new}=5\times 10^{-2}$ eV$^{2}$. For large $\Delta m^{2}_{new}$ values, the 90\% C.L. upper limit on \sin22tnew reaches $7.2 \times 10^{-3}$. This result is still affected by the statistical underfluctuation, the sensitivity corresponding to the analysed statistics being $10.4 \times 10^{-3}$. A Bayesian statistical treatment has therefore been adopted for determining the upper limit.

Various improvements are expected for the future. The statistics will be increased by a factor of 3.4 by completing the analysis of the collected data. The reconstructed energy resolution will be improved when the calorimetric measurement in the TT will be complemented by following the hadron tracks and the electron showers in the downstream bricks. 

With the increase in sample size and the improvements in the analysis, the effect of a possible statistical underfluctuation of the background will be reduced and OPERA should then be able to access the parameter region comparable to its sensitivity below \sin22tnew = 5.0$\times 10^{-3}$.

\acknowledgments

We thank CERN for the successful operation of the CNGS facility and INFN for the continuous
support given to the experiment during the construction, installation and data-taking
phases through its LNGS laboratory. We warmly acknowledge funding from our national
agencies: Fonds de la Recherche Scientifique-FNRS and Institut Interuniversitaire des
Sciences Nucl\'eaires for Belgium, MoSES for Croatia, CNRS and IN2P3 for France, BMBF for
Germany, INFN for Italy, JSPS (Japan Society for the Promotion of Science), MEXT (Ministry
of Education, Culture, Sports, Science and Technology), QFPU (Global COE programme of
Nagoya University, Quest for Fundamental Principles in the Universe supported by JSPS and
MEXT) and Promotion and Mutual Aid Corporation for Private Schools of Japan for Japan,
SNF, the University of Bern and ETH Zurich for Switzerland, 
the Russian Foundation for Basic Research (grant no. 09-02-00300 a, no.12-02-12142-ofi-m), 
the Programs of the Presidium of the Russian Academy
of Sciences Neutrino physics and Experimental and theoretical researches of fundamental
interactions connected with work on the accelerator of CERN, the Programs of Support of
Leading Schools (grant no. 3517.2010.2), and the Ministry of Education and Science of the
Russian Federation for Russia, the Korea Research Foundation Grant (KRF-2008-313-C00201)
for Korea and TUBITAK, the Scientific and Technological Research Council of Turkey, for
Turkey. We are also indebted to INFN for providing fellowships and grants to non-Italian
researchers. We thank the IN2P3 Computing Centre (CC-IN2P3) for providing computing
resources for the analysis and hosting the central database for the OPERA experiment. We are
indebted to our technical collaborators for the excellent quality of their work over many years
of design, prototyping and construction of the detector and of its facilities.



\end{document}